\documentstyle[aps]{revtex}

\begin{document}
\title{{\bf Light-front time picture of few-body systems}}
\author{T. Frederico$^{a}$, J. H. O. Sales$^{b}$, B. V. Carlson$^{a}$ and \and P. U.
Sauer$^{c}$.}
\address{$^{a}$Dep. F\'{\i}sica Instituto Tecnol\'{o}gico de Aeron\'{a}utica, \\
Centro T\'{e}cnico Aeroespacial, 12.228-900 \\
S\~{a}o Jos\'{e} dos Campos, S\~{a}o Paulo, Brazil.\\
$^{b}$Instituto de F\'{\i}sica Te\'{o}rica-UNESP, 01405-900\\
S\~{a}o Paulo, Brazil.\\
$^{c}$Institute for Theoretical Physics, University Hannover \\
D-30167 Hannover, Germany}
\maketitle

\begin{abstract}
The projection of the four-dimensional two-body Bethe-Salpeter equation at
the light-front hypersurface is discussed. Elimination of the relative time
is performed using a quasi-potential expansion, which is shown to be
equivalent to an infinite set of coupled equations for the Green's functions
of the different light-front Fock-states.
\end{abstract}


\section{INTRODUCTION}

The field theoretical Bethe-Salpeter equation (BSE) in the ladder
approximation can be used as starting point to model bound-states of
relativistic few-body systems. The light-front time dynamics \cite{8} leads
to the representation of the two-body BSE in which an effective mass squared
operator arises and has as an eigenfunction the lowest Fock-state component
(valence) of the wave-function \cite{j1,j2}. The BSE is equivalent to an
infinite set of coupled equations for Green's functions with different
number of particles, which also allows a consistent truncation of the
light-front Fock-space. This set of coupled equations comes from a
quasi-potential expansion of the BSE on the light-front \cite{j1,j2,2}. The
relative light-front time between the particles is eliminated in favor of a
global time.

The two-body bound-state light-front equation with the lowest order kernel 
\cite{4} (truncation of the virtual intermediate states up to
three-particles) was applied to study bosonic \cite{5} and fermionic \cite
{6,brodsky} systems. Nowadays, there is a renewed interest in studying the
structure of few-body bound-states at the light-front time $x^+=t+z=0$, due
to the possibility of probing hadrons and few-nucleon systems at short
distances \cite{7}.

We perform the quasi-potential reduction of two-body BSE's and present the
coupled set of equations for the light-front Green's functions for bosonic
and fermionic models, with the interaction Lagrangian respectively given by, 
${\cal L}^B_I=g_S\phi _1^{\dagger }\phi _1\sigma +g_S\phi _2^{\dagger }\phi
_2\sigma$ and ${\cal L}_{I}^F=g_{S}\overline{\Psi }\Psi\sigma$, where $%
\phi_1 $, $\phi_2$ and $\sigma$ are the bosonic fields and $\Psi$ is the
fermion field in the Yukawa model.

\section{GLOBAL $x^+$-TIME AND QUASI-POTENTIAL REDUCTION}

The starting point is the BSE in relativistic field theory for the two-body
transition matrix, $T(K)$, with total four-momentum $K$ and two-body
irreducible interaction $V(K)$: 
\begin{eqnarray}
T(K) = V(K)+V(K)G_0(K)T(K) ~.  \label{eq2}
\end{eqnarray}
Here, we neglect self-energy terms in the disconnected Green's function.

We begin by discussing the elimination of the relative $x^+$-time in (\ref
{eq2}) for a two-boson system \cite{j1}. The free Green's function expressed
in terms of {\it light-front} kinematical momenta, e.g., $%
k_i=(k^-_i:=k^0_i-k^3_i\ , \ k^+_i:=k^0_i+k^3_i \ , \ \vec k_{i\perp})$, is: 
\begin{eqnarray}
\left\langle k_1^{\prime -}\right| G_0(K)\left|k^-_1\right\rangle =-\frac1{%
2\pi} \frac{\delta \left(k^{\prime -}_1-k^{-}_1\right)}{\widehat k^+_1 (K^+-%
\widehat k^+_1) \left(k_1^--\frac{\widehat{\vec k}^2_{1\perp}+m_1^2-io}{%
\widehat k^+_1}\right) \left(K^--k_1^--\frac{\widehat{\vec k}%
^2_{2\perp}+m_2^2-io}{K^+-\widehat k^+_1} \right)}~,  \label{eq3}
\end{eqnarray}
where operators are denoted by a "hat" and $\widehat k^-_{ion}=(\widehat{%
\vec k}_{i\perp}^2+m^2)/ \widehat k^+_i$ ($i=1,2$). The kinematical momentum
states, $\langle x^-_i \vec x_{i\perp}\left| k^+_i\vec k_{i\perp}\right%
\rangle= e^{-\imath(\frac12 k^+_ix^-_i-\vec k_{i\perp} . \vec x_{i\perp})}$,
are eigenfunctions of the free energy operator $\widehat k^-_{ion} $. The
completeness relation and the normalization are $\int \frac{dk^+d^2k_\perp}{%
2(2\pi)^3} \left. |k^+\vec k_\perp\rangle \langle k^+\vec k _\perp \right. |
=1$ and $\langle k^{\prime +}\vec k^\prime_\perp |k^+\vec k_\perp
\rangle=2(2\pi)^3\delta(k^{\prime +}-k^+)\delta(\vec k^{\prime}_\perp-\vec k%
_\perp)$.

The integrations over $k^-_1$ and $k^{\prime-}_1$ eliminate the relative
time: 
\begin{eqnarray}
g_0(K)=|G_0(K)|:= \int dk^{\prime -}_1 dk^{ -}_1 \left\langle k_1^{\prime
-}\right| G_0(K)\left|k^-_1\right\rangle \ =i\widehat\Omega^{-1}g_0^{(2)}(K)%
\widehat\Omega^{-1} \ ,  \label{eq4}
\end{eqnarray}
and the Fourier transform in $K^-$ gives the free propagator in the $x^+$%
-time. Here, $K^+ > 0$ is used without any loss of generality. The phase
space operator is $\widehat\Omega:=\sqrt{\widehat k_1^+(K^+-\widehat k^+_1)}$%
. The vertical bar $|$ on the left of the Green's function represents
integration on $k^-_1$ in the bra-state, the bar on the right in the
ket-state. The free two-body light-front Green's function is a particular
case of the light-front Green's function for $N$ particles: 
\begin{eqnarray}
g^{(N)}_0(K)= \left[\prod _{j=1}^N \theta (\widehat k_j^{+})\theta (K^{+}-%
\widehat k_j^{+})\right]\left( K^{-}-\widehat K_0^{(N)-}+io \right)^{-1} ,
\end{eqnarray}
where $\widehat K_0^{(N)-}=\sum_{j=1}^N\widehat{k}_{j\ on}^-$ is the free
light-front Hamiltonian.

For fermions, the Dirac propagator is written as $({\rlap\slash
k} -m+io)^{-1}= ( {\rlap\slash k}_{on} +m)/(k^2-m^2+io)+ \gamma^+/(2k^+)$,
to separate the last term which is instantaneous in the $x^+$-time.
Therefore, the free two-fermion propagator can be written as: 
\begin{eqnarray}
G^F_0(K) = \left({\rlap\slash k}_1 +m_1\right) \left({\rlap\slash k}_2
+m_2\right) G_0(K) = \Delta G_0^F(K) +{\overline G}^F_0(K)\ ,
\end{eqnarray}
where $\Delta G_0^F(K)$ contains the instantaneous terms and the global
light-front time propagator of the two fermion system is ${\overline {G}%
^F_0(K)} =\left( {\rlap\slash k}_{1on} +m\right) \left({\rlap\slash k}_{2on}
+ m\right)G_0(K) $. Here, we only define the basis states of spinorial
functions in the positive energy sector as $\langle x^{-}\vec{x}_{\perp
}\left| k^{+}\vec{k}_{\perp }\ s\right\rangle =e^{-i(\frac{1}{2}k^{+}x^{-}-%
\vec{k}_{\perp }.\vec{x}_{\perp })}u(k,s)$. The states form a complete basis
for positive energy spinor functions, 
\begin{equation}
\sum_{s}\int \frac{dk^{+}d^{2}k_{\perp }}{2(2\pi )^{3}}\left| k^{+}\vec{k}%
_{\perp }\ s\right\rangle \left\langle k^{+}\vec{k}_{\perp }s\right| \gamma
^{0}=\mbox{\boldmath$1$}~, ~~ u(k,s)=\frac{\left( {\rlap\slash k}%
_{on}+m\right) }{\sqrt{2k^{+}2m}}\gamma ^{+}\gamma ^{0}\left( 
\begin{array}{c}
\chi _{s} \\ 
0
\end{array}
\right);
\end{equation}
where $\chi _{s}$ is the two-component Pauli spinor and $u(k,s)$ is the
light-front spinor \cite{brodsky}. We point out that one could equally well
study particle-antiparticle systems or others.

Our aim here is to discuss the integral equation for the interacting
light-front two-body propagator, $g(K)$, which is derived from 
\begin{eqnarray}
G(K) = \overline G_0(K)+ \overline G_0(K)T(K)\overline G_0(K) ,  \label{eq5}
\end{eqnarray}
with $g(K)\equiv|G(K)|$. For a two-boson system $\overline G^B_0(K)\equiv
G_0(K)$. We use the quasi-potential reduction of Ref. \cite{wolja} to
perform the projection to $x^+=0$. For that purpose an auxiliary Green's
function $\widetilde{G}_0(K)$ is introduced as 
\begin{eqnarray}
\widetilde G_0(K):= {\overline G}_0(K)| g_0^{-1}(K) |\overline{G}_0(K) \ ,
\label{eq6}
\end{eqnarray}
which allows one to separate the light-front two-particle propagation in the
intermediate states \cite{j1,j2}. In our discussion, the operator $%
g_0^{-1}(K)$ for a two-fermion system is defined in the subspace of positive
energy spinors.

The transition matrix is the solution of 
\begin{eqnarray}
&& T(K) = W(K)+W(K)\widetilde{G}_0(K)T(K) ,  \label{eq7} \\
&& W(K)=V(K)+V(K)[G_0(K)-\widetilde{G}_0(K)]W(K) ,  \label{eq8}
\end{eqnarray}
according to Ref. \cite{wolja}. The auxiliary three-dimensional transition
matrix 
\begin{eqnarray}
t(K)=g_0(K)^{-1}|{\overline G}_0(K)T(K)\overline{G}_0(K)|g_0(K)^{-1}~,
\label{eq9}
\end{eqnarray}
gives the scattering amplitude for two-particles, which is identical to the
one obtained directly from $T$ (see discussion in \cite{j1,j2} ). The
auxiliary transition matrix satisfies 
\begin{eqnarray}
t(K)=w(K)+w(K)g_0(K)t(K),~w(K)=g_0(K)^{-1}|\overline G_0(K)W(K)\overline G%
_0(K)|g_0(K)^{-1} ~;  \label{eq11}
\end{eqnarray}
which can be derived using Eqs. (\ref{eq7}) and (\ref{eq9}).

\section{HIERARCHY EQUATIONS FOR LF GREEN'S FUNCTIONS}

The two-body light-front propagator satisfies $g^{(2)}(K)=
g^{(2)}_0(K)+g^{(2)}_0(K) \nu (K) g^{(2)}(K)$ where $g^{(2)}(K)\equiv -i%
\widehat\Omega g(K)\widehat\Omega$, $\nu (K)=i \widehat\Omega^{-1} w(K)%
\widehat\Omega^{-1}$, which is derived from Eqs. (\ref{eq5}) and (\ref{eq11}%
). The poles of $g^{(2)}(K)$ are at the positions of the bound states, and
from its residue one can obtain the lowest Fock component of the wave
function. In addition, the light-front description fully retrieves the
covariant one \cite{j1}.

The first order contribution to $\nu(K)$ obtained from Eqs. (\ref{eq8}) and (%
\ref{eq11}) is 
\begin{eqnarray}
\nu^{(2)}(K)=i \left[\widehat\Omega g_0(K)\right]^{-1}|G_0(K)V(K)G_0(K)| %
\left[g_0(K)\widehat\Omega\right]^{-1}\ .  \label{eq14}
\end{eqnarray}
The second order term in the quasi-potential expansion given by 
\begin{eqnarray}
\nu^{(4)}(K)&=&i \left[\widehat\Omega g_0(K)\right]^{-1}|G_0(K)V(K)
G_0(K)V(K)G_0(K)|\left[g_0(K)\widehat\Omega\right]^{-1}  \nonumber \\
&-&i\left[\widehat\Omega g_0(K)\right]^{-1}|G_0(K)V(K)\widetilde G%
_0(K)V(K)G_0(K)|\left[g_0(K)\widehat\Omega\right]^{-1} \ ,  \label{eq16}
\end{eqnarray}
is two-body irreducible, due to the subtraction of the second term in the
r.h.s. of Eq. (\ref{eq16}).

We restrict our discussion to the ladder approximation of the BSE. Below, we
explicit the content of the operator $\nu^{(n)}$ in the light-front
Fock-space. For the chosen bosonic Lagrangian model, the interaction
operator acting between Fock-states differing by one quantum $\sigma$, has
matrix element given by 
\begin{eqnarray}
\langle q k_\sigma |v|k\rangle= -2(2\pi)^3\delta (q+k_\sigma-k) \frac{g_S}{%
\sqrt{q^+k^+_\sigma k^+}} \theta (k^+_\sigma)~ ,  \label{eq18}
\end{eqnarray}
where the delta function expresses the kinematical momentum conservation.
The effective interaction $\nu (K)$ up to second order in $v$ is: 
\begin{eqnarray}
\nu (K)\approx \nu ^{(2)}(K)+\nu^{(4)}(K)=vg_0^{(3)}(K)v
+vg_0^{(3)}(K)vg_0^{(4)}(K)vg_0^{(3)}(K)v \ ,  \label{eq21}
\end{eqnarray}
where, from Eq.(\ref{eq14}), $\nu^{(2)}$ is $vg_0^{(3)}(K)v$. The
contribution of four-body intermediate states to the three-body propagation
arises from Eq.(\ref{eq16}). In fact, $\nu(K)$ is equal to $vg^{(3)}(K)v$,
which is found when the expansion is performed to all orders. From Eq.(\ref
{eq21}), one sees that $g^{(3)}$ is coupled to the four-body Green's
function, which should be coupled to the five-body one, and so on. By
generalization, one can construct a hierarchy of coupled equations for the
Green's functions: 
\begin{eqnarray}
&&g^{(2)}(K)=g^{(2)}_0(K) +g^{(2)}_0(K)v g^{(3)}(K)vg^{(2)}(K)~, \ . \ . \ .
\nonumber \\
&& g^{(N)}(K)=g^{(N)}_0(K) +g^{(N)}_0(K)v g^{(N+1)}(K)vg^{(N)}(K)~, \ . \ .
\ . \   \label{eq22}
\end{eqnarray}

In the Yukawa model for fermions, the interaction operator acting between
Fock-states differing by zero, one and two $\sigma$'s, has matrix elements
given by 
\begin{eqnarray}
&&\langle (q,s^\prime) k_\sigma |v|(k,s)\rangle= -2m(2\pi)^3\delta
(q+k_\sigma-k) \frac{g_S}{\sqrt{q^+k^+_\sigma k^+}} \theta (k^+_\sigma){%
\overline u}(q,s^\prime)u(k,s)~,  \label{eq18a} \\
&& \langle (q,s^\prime)k^\prime_\sigma|v|(k,s)k_\sigma\rangle=
-2(2\pi)^3\delta (q+k^\prime_\sigma-k-k_\sigma)\delta_{s^\prime s} \frac{%
g^2_S}{\sqrt{k^{\prime +}_\sigma k^+_\sigma}} {\frac{\theta (k^{\prime
+}_\sigma)\theta (k^+_\sigma) }{k^++k_\sigma^+}} ~,  \label{eq18b} \\
&&\langle (q,s^\prime) k^\prime_\sigma k_\sigma|v|(k,s) \rangle=
-2(2\pi)^3\delta (q+k^\prime_\sigma+k_\sigma-k )\delta_{s^\prime s} \frac{%
g^2_S}{\sqrt{k^{\prime +}_\sigma k^+_\sigma}} {\frac{\theta (k^{\prime
+}_\sigma)\theta (k^+_\sigma) }{k^+-k_\sigma^+}} ~.  \label{eq18c}
\end{eqnarray}
The instantaneous terms in the two-fermion propagator give origin to Eqs. (%
\ref{eq18b}) and (\ref{eq18c}). The expansion of $\nu(K)$ has terms that
couple sectors of the Fock-space that differ at most by two sigma's \cite{j2}%
. Therefore, the coupled set of Green's function found for bosons has to be
extended for the Yukawa model as: 
\begin{eqnarray}
&&g^{(2)}(K)=g^{(2)}_0(K)+g^{(2)}_0(K)v\left[ g^{(3)}(K)+g^{(4)}(K)
+g^{(3)}(K)vg^{(4)}(K) \right.  \nonumber \\
&&+\left. g^{(4)}(K)vg^{(3)}(K)\right]vg^{(2)}(K)~, \ . \ . \ .  \nonumber \\
&&g^{(N)}(K)=g^{(N)}_0(K) +g^{(N)}_0(K)v \left[ g^{(N+1)}(K)+g^{(N+2)}(K) +
g^{(N+1)}(K)vg^{(N+2)}(K) \right.  \nonumber \\
&&+\left.g^{(N+2)}(K)vg^{(N+1)}(K)\right]vg^{(N)}(K)~, \ . \ . \ .
\label{hfer}
\end{eqnarray}

The quasi-potential expansion differs from the truncation in the light-front
Fock space. Taking the two-boson system as an example and restricting the
intermediate state propagation up to four-particles, we find that $%
g^{(2)}(K)=g^{(2)}_0(K) +g^{(2)}_0(K)v g^{(3)}(K)vg^{(2)}(K)$ and $%
g^{(3)}(K)=g^{(3)}_0(K) +g^{(3)}_0(K)v g^{(4)}_0(K)vg^{(3)}(K)$; which goes
beyond the quasi-potential expansion in second order, where, instead of $%
g^{(3)}(K)$, one has $g^{(3)}_0(K)$ (c.f. Eq.(\ref{eq21})).

\section{SUMMARY}

The set of coupled equations, (\ref{eq22}) and (\ref{hfer}), contains all
two-body irreducible diagrams with the exception of those including closed
loops in $\Phi_1$ and $\Phi_2$ or in $\Psi$ and as well part of the
cross-ladder diagrams. The light-front Fock states contains two bosons $%
\Phi_1$ and $\Phi_2$ or two fermions with any number of bosons $\sigma $.
This excludes the complete representation of the crossed ladder diagrams.
The hierarchy set of equations also resembles the iterated resolvent method 
\cite{brodsky}. The covariant ladder restricts the kernel of the hierarchy
equations, i.e., crossed terms and self energies are not included in Eqs. (%
\ref{eq22}) and (\ref{hfer}). This set of equations can also be useful in
the study of renormalization in the light-front, as has been discussed for
fermions in the context of the quasi-potential expansion \cite{j2}.

In summary, we have presented the general framework for constructing the
light-front two-body Green's function using a quasi-potential reduction of
the Bethe-Salpeter equation. We have found a coupled set of equations which
gives the two-body propagator in several cases, including the covariant
ladder approximation. Finally, we have discussed within our scheme the
truncation of the light-front Fock-space. Renormalization and local gauge
invariance in this framework is a challenge open for future work.

We thank the Brazilian agencies FAPESP and CNPq for partial support.

\end{document}